\documentclass[12pt]{elsarticle}

\makeatletter
\def\ps@pprintTitle{%
 \let\@oddhead\@empty
 \let\@evenhead\@empty
 \def\@oddfoot{\centerline{\thepage}}%
 \let\@evenfoot\@oddfoot}
\makeatother

\usepackage{ifpdf}
\usepackage{cite}

\usepackage{amsmath}
\usepackage{algorithmic}
\usepackage{array}
\usepackage{fixltx2e}
\usepackage{stfloats}
\usepackage{url}
\usepackage{amssymb} 
\usepackage{multicol}
\usepackage{listings}
\usepackage{rotating} 
\usepackage{enumitem}
\usepackage{color}
\usepackage{multirow}
\usepackage[acronym]{glossaries}
\usepackage{graphicx}
\usepackage{float}
\usepackage[export]{adjustbox}
\usepackage{times}
\usepackage{geometry}
\usepackage{blindtext}
\usepackage{setspace}
\usepackage{tocloft}
\usepackage[center]{caption}
\usepackage{tabu}
\journal{Pre-print}

\begin{document}

\begin{frontmatter}


\title{Assistive Completion of Agrammatic Aphasic Sentences: A Transfer Learning Approach using Neurolinguistics-based Synthetic Dataset}



\author{Rohit Misra}
\author{Sapna S Mishra}
\author{Tapan K. Gandhi}
\address{ Department of Electrical Engineering,  
Indian Institute of Technology Delhi,\\
Hauz Khas, New Delhi, 110016, India
}

\begin{abstract}

Damage to the inferior frontal gyrus (Broca’s area) can cause agrammatic aphasia wherein patients, although able to comprehend, lack the ability to form complete sentences. This inability leads to communication gaps which cause difficulties in their daily lives. The usage of assistive devices can help in mitigating these issues and enable the patients to communicate effectively. However, due to lack of large scale studies of linguistic deficits in aphasia, research on such assistive technology is relatively limited. In this work, we present two contributions that aim to re-initiate research and development in this field. Firstly, we propose a model that uses linguistic features from small scale studies on aphasia patients and generates large scale datasets of synthetic aphasic utterances from grammatically correct datasets. We show that the mean length of utterance, the noun/verb ratio, and the simple/complex sentence ratio of our synthetic datasets correspond to the reported features of aphasic speech. Further, we demonstrate how the synthetic datasets may be utilized to develop assistive devices for aphasia patients. The pre-trained T5 transformer is fine-tuned using the generated dataset to suggest 5 corrected sentences given an aphasic utterance as input. We evaluate the efficacy of the T5 model using the BLEU and cosine semantic similarity scores. Affirming results with BLEU score of 0.827/1.00 and semantic similarity of 0.904/1.00 were obtained. These results provide a strong foundation for the concept that a synthetic dataset based on small scale studies on aphasia can be used to develop effective assistive technology.
\end{abstract}


\end{frontmatter}


\section{Introduction}
\label{S:1}
Language is one of the vital abilities that differentiates human beings from other forms of life. Humans have evolved to communicate in highly structured forms with dense vocabularies in various languages. Studies show that areas in the brain have specialised for comprehending language using both auditory and visual inputs. A closely linked system also operates in the brain which is responsible for generation of structured sentences and speech from abstract thoughts (\citet{friederici2011brain}). The region of brain responsible for sentence formation and speech production is called the Broca's area. Damage to this region due to stroke or injury results in a disorder termed Broca's Aphasia (\citet{acharya2021broca, udden2017broca}). Broca's aphasia is a neurological impairment wherein the patient is unable to generate fully formed sentences or produce corresponding speech to converse. The comprehension abilities and intelligence of the patient are not affected in this form of aphasia (\citet{acharya2021broca}). This leads to a breakdown between the person's thoughts and language capabilities. Patients often feel that they know what they wish to say but are unable to produce the words. 

Rehabilitation and assistance of patients with such aphasia is a matter of concern. Various rehabilitation camps and courses with regular, step-wise training have been  successful in helping the patients regain command over spoken languages (\citet{rose2014aphasia}, \citet{van2011aphasia}, \citet{sarno1993aphasia}). However, these programs take long periods of time ranging from months to years. Therefore, in order to alleviate the immediate impact of loss of verbal communication in the lives of aphasics, assistive technology may be employed. Examples of assistive technology include, written text assistance, interactive devices with automatic word suggestions, or ones that complete the agrammatic sentences uttered by the aphasia patients (\citet{mcgrenere2002insights}, \citet{wallace2021reading}). Development of such assistive devices requires detailed studies and characterisation of the speech deficits observed in aphasics. Neurological (\citet{friederici2011brain}) as well as linguistic (\citet{tetzloff2018quantitative}, \citet{thompson1995analysis}) analyses shall be crucial in enhancing our understanding of the language processing mechanisms in the brain and related disorders.

Functional linguistic analysis of utterances made by Broca's aphasia patients has aided neurolinguists to objectively describe the agrammatic nature of aphasic speech (\citet{saffran1989quantitative}). Additionally, major crucial developments made in Natural Language Processing (NLP) have enabled computers to process human languages using deep learning models (\citet{vaswani2017attention}, \citet{brown2020language}, \citet{hu2020xtreme}, \citet{yang2019xlnet}, \citet{devlin2018bert}). Various large deep learning models have been proposed and trained to perform tasks like semantic analysis (\citet{salloum2020survey}), summarization (\citet{devlin2018bert, miller2019leveraging, grail2021globalizing}), translation (\citet{hu2020xtreme}), etc. Text-to-Text transformer models provide a gateway into interactive applications of NLP (\citet{raffel2020exploring}). Training the deep learning models for these tasks requires large datasets. However, as the number of patients of brain injury or stroke are limited, it is difficult to obtain large collections of sentences that were uttered by aphasics. Moreover, as Broca's aphasia is a speech disorder, constructing such a dataset would require tremendous efforts in recording and transcribing. Due to this lack of large datasets, NLP research focused on speech disorders like aphasia and dyslexia is very limited (\citet{themistocleous2021part}).

In this work, we propose a novel concept of an assistive device for patients of Broca's aphasia that advances towards the amalgamation of neurolinguistics and NLP research. Concerning the problem of unavailability of large datasets of aphasic speech, we propose the generation of a synthetic dataset that comprises agrammatic sentences which functionally match the linguistic characteristics of utterances of aphasia patients. We present a model that generates synthetic aphasic utterances given a corpus of grammatically correct sentences. Small scale quantitative studies on the linguistic characterisation of aphasic utterances have been used as a foundation for the model. It is proposed to use this synthetic dataset to fine tune a transfer learning model for a sentence completion task. The trained model shall take the agrammatic aphasic text as input and suggest possible reconstructed sentences.

\section{Linguistic Characterization and Proposed Model}
\begin{figure}
    \centering
    \includegraphics[width = 0.65\textwidth]{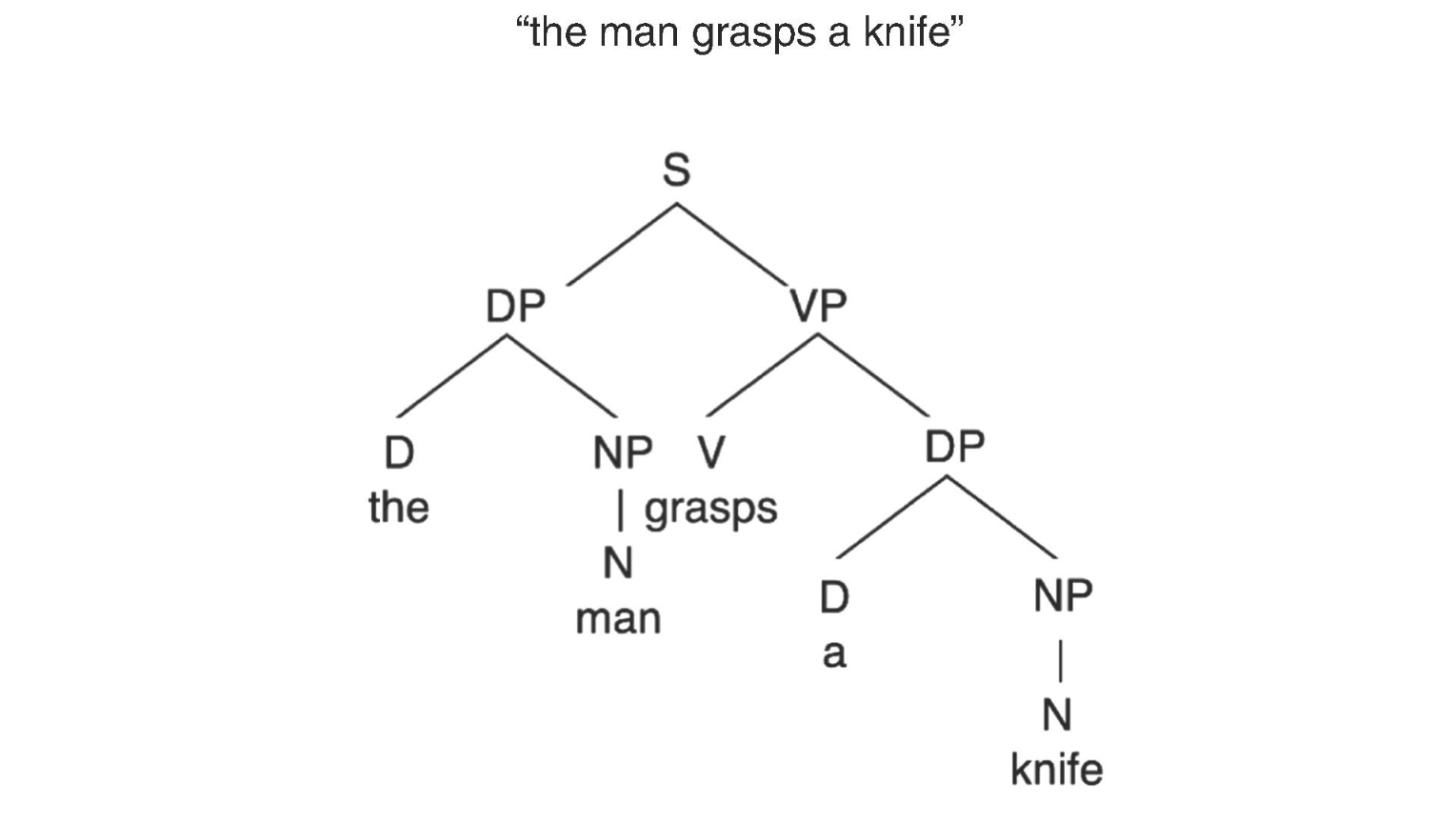}
    \caption{Tree parsing of a sentence to show component phrases (\citet{zaccarella2021language})}
    \label{fig:my_label}
\end{figure}

Linguistics encompasses the study of various aspects of language like syntax, semantics, functional analysis, modelling, etc. In the English language, commonly used parts of speech include nouns, pronouns, verbs, adjectives, etc. Words from different parts of speech are combined grammatically to form sentences. A sentence can be functionally segregated into various phrases. Primary phrases in a sentence may be the subject noun phrase (NP), verb phrase (VP), object noun phrase (NP) or the determiner phrase (DP). Such phrases contain the corresponding root words along with appropriate modifiers like articles, adjectives, adverbs, etc. Primary phrases may be further divided into similar sub-phrases until only single words are obtained depending on the complexity of the sentence (Figure \ref{fig:my_label}). A complex sentence contains verbs with multiple arguments leading to more noun phrases in the sentence per verb phrase. In this section we present the functional features of aphasic utterances followed by the proposed model for generating synthetic aphasic speech.

\subsection{Functional Features of Aphasic Speech}

Broca’s Aphasia is also known as Agrammatism or Agrammatic Aphasia  as the sentence formation ability is weakened in the brain. Studies of interactions with aphasic patients (\citet{saffran1989quantitative}, \citet{gleason1975retrieval}, \citet{grodzinsky1984syntactic}, \citet{tetzloff2018quantitative}, \citet{kean1977linguistic, caplan1985syntactic}) have shown that the following phenomena are common in aphasic sentence production (\citet{berndt2002production}, \citet{avrutin2001linguistics}, \citet{gleason1980narrative}, \citet{meyerson1972transformational}, \citet{berndt1996comprehension}). 

\begin{itemize}
    \item \textbf{Determiner Error:} Agrammatic speech is commonly devoid of determiners like “a”, “an”, “the”, etc.
    \item \textbf{Omission Errors:} A characteristic feature of agrammatic speech is frequent omission of functional categories, such as determiners, tense, and complementizers.
    \item \textbf{Subject Drop:} Omission of subject NPs in the speech of Broca’s aphasics is well documented. This is similar to “Diary Speech” in English (\citet{haegeman2013syntax}).
    \item \textbf{Verb Deficit:} Individuals are able to produce one-argument verbs better than two- or three-argument verbs. Agrammatic patients often delete obligatory verb arguments in sentences.
\end{itemize}

\subsection{Linguistics-based Model of Aphasic Sentence Generation}
Based on the aforementioned functional linguistic features of aphasic utterances, we propose a model that converts a grammatically correct sentence to a possibly aphasic sentence. As an input dataset of correct sentences, the C4 dataset (\citet{raffel2020exploring}) containing a large set of sentences extracted from websites is used.

\begin{figure}[!h]
    \centering
    \includegraphics[width = \textwidth]{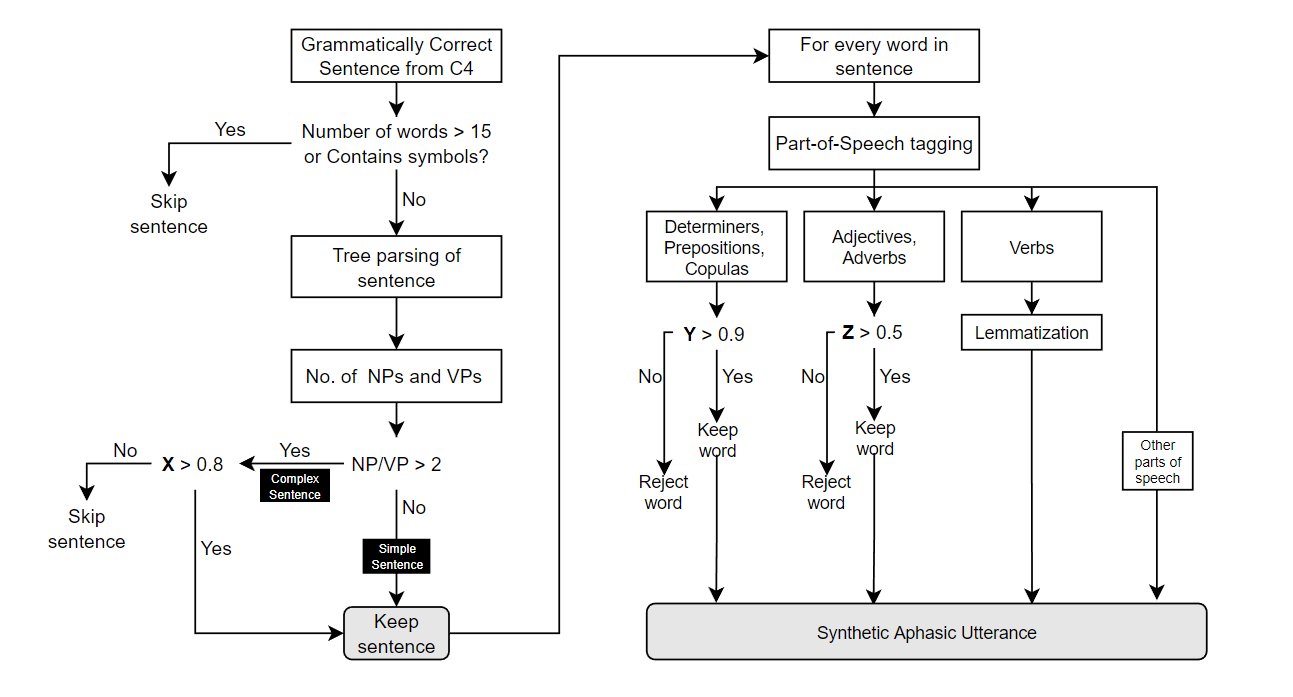}
    \caption{Flow chart of the proposed model for generation of synthetic aphasic utterance. Here \textbf{X}, \textbf{Y}, and \textbf{Z} are continuous random variables uniformly distributed in [0,1]}
    \label{flowchart}
\end{figure}

From the C4 dataset, any sentences that are longer that fifteen words are rejected to mimic the average small length of utterance in aphasia. Sentences containing "?" or other symbols are also removed from the dataset because of a lack of studies that quantitatively characterise these aspects of agrammatic speech. Further, to emulate the absence of helping words like determiners, prepositions, or copulas, these  parts-of-speech are removed from sentences with a probability of 90\%. Omission errors in aphasic speech are incorporated in the model by rejecting adjectives and adverbs with a 50\% probability. The verbs present in the sentences are also converted to their root form by the process of \textit{lemmatization} (\citet{manning2014stanford}). According to quantitative linguistic studies (\citet{tetzloff2018quantitative}, \citet{thompson1995analysis}), the probability of utterance of sentences containing complex verbs with more than two obligatory arguments is negligible in aphasia patients. So, to implement this, we use the ratio of the total number of noun phrases to the total verb phrases (NP/VP) as a measure of sentence complexity. Sentences with NP/VP ratio greater than two were considered complex and were rejected from the dataset with a probability of 80\%. Performing these operations on the sentences from the C4 generated pairs of grammatically correct and corresponding modeled aphasic utterances. Figure \ref{flowchart} shows the graphical representation of the model. 

\section{Generation of Synthetic Dataset}
 Advanced text-to-text deep learning models can be trained on massive datasets for applications in semantic analysis, summarization, error correction, etc. \citet{stahlberg2021synthetic} have proposed the idea of generation of synthetic datasets for tasks like Grammatical Error Correction (GEC) for which tagged datasets are scarce. They have attempted to match the frequency and type of synthetic errors to normal human conversational grammatical errors. In addition, Masked Language Models (MLMs) are another training paradigm proposed by \citet{devlin2018bert} for bidirectional text-to-text transformers. In the MLM, certain subset of words in a given sentence are masked and a transformer is trained to predict those masked words based on other words in that sentence. This type of task adds complexity to regular language model task, and it can help boost performance.

The MLM training procedure follows a random masking strategy. However, the agrammatic errors in Broca’s Aphasia are not random in nature. As reported in linguistic studies on aphasic patients (detailed in the previous section), the omission errors are dependent on the parts of speech and also the complexity of the sentence.  Inspired by the aforementioned works, we use the generated synthetic dataset of aphasic sentences to fine tune a text-to-text transformer for a sentence completion task. The training set used here contains 20,000 pairs of aphasic and corresponding correct sentences. The testing set contains 10,000 such pairs. 

\begin{figure}[!h]
     \centering
     \includegraphics[width = \textwidth]{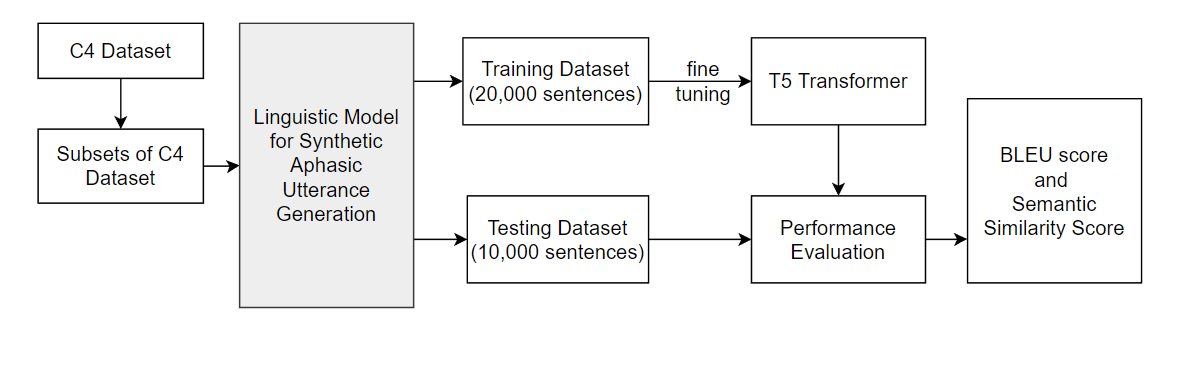}
     \caption{Proposed model for dataset generation and fine tuning}
     \label{procedure}
 \end{figure}

In present work, the Text-to-Text Transfer Transformer (T5) (\citet{2020t5}) has been utilized. The T5 is a text-to-text deep learning model that is pre-trained on the massive C4 dataset for general purpose NLP tasks. It can be fine tuned using a smaller dataset for any specific text-to-text application. In this work, the T5 is trained to complete a sentence given the aphasic input. The synthetic dataset is used to fine tune the T5 model. The task prefix used for T5 is: “\textit{Complete this sentence:}”. The training paradigm uses the Adam optimizer with a learning rate of $4 \times 10^{-5}$. The training data was utilized in 16 batches and the model was trained for 3 epochs. The trained model yields 5 possible reconstructed sentences given an aphasic sentence. an overview of the proposed procedure for dataset generation and fine-tuning is shown in Figure \ref{procedure}

\section{Results}
The train and test datasets were generated using the C4 corpus. In order to functionally compare the synthesized agrammatic sentences with aphasic utterances, we use certain metrics as mentioned below.

\begin{itemize}
    \item \textbf{Mean Length of Utterance}: The average length of the sentences uttered by aphasia patients is significantly lower than healthy speakers. 
    \item \textbf{Noun-to-Verb Ratio}: The noun-to-verb ratio in aphasic speech is a reflection of the number of grammatically correct subject-object relations in the utterances. It is also an indirect measure of the complexity of the verbs used by the speaker as complex verbs require more arguments, and thus more nouns.
    \item \textbf{Simple-to-Complex Sentence Ratio}: Aphasic speech is characterized by a significantly lower number of complex sentences that contain verbs with more than two arguments (\citet{thompson1995analysis}).   
\end{itemize}

These metrics, obtained for the training and testing synthetic datasets are reported in Table \ref{Table:datasets}. The same metrics were also calculated for a subset of the C4 dataset (of size 20,000) to analyse the difference between grammatically correct sentences and synthesized aphasic sentences.  

\begin{table}[!h]
\caption{Linguistic features of C4 and synthetic datasets}
\label{Table:datasets}
\resizebox{\textwidth}{!}{%
\begin{tabular}{|l|l|l|l|}
\hline
\textbf{Metrics}                                    & \textbf{Subset of C4 Dataset} & \textbf{Synthesized Training Dataset} & \textbf{Synthesized Testing Dataset} \\ \hline
\textbf{Number of sentences}                 & 20,000                        & 20,000                                & 10,000                               \\
\textbf{Mean length of utterance (L)}        & 18.97 words                         & 6.99 words                            & 7.01 words                           \\
\textbf{Noun/Verb ratio (N/V)}               & 2.96                          & 2.41                                  & 2.44                                 \\
\textbf{Simple/Complex sentence ratio (S/C)} & 1.70                          & 9.08                                  & 9.11                                 \\ \hline
\end{tabular}%
}
\end{table}

To evaluate the performance of the sentence completion model, we compare the degree of similarity between the original gramatically correct sentences from the C4 dataset and the corrected sentences predicted by the model. Here, the BiLingual Evaluation Understudy (BLEU) score is used as a metric to evaluate the model performance (\citet{papineni2002bleu}). The performance of the model on the testing dataset is also evaluated using cosine semantic similarity of predicted and intended sentences (\citet{gomaa2013survey}). Among the 5 corrected sentences suggested by the model, the suggestion with the highest score for the given sentence was considered for mean score calculation. Overall, the sentence completion model showed affirming performance with average BLEU score = 0.827/1.00 and semantic similarity = 0.904/1.00.

\section{Discussion}

With the present work, we put forth a concept that aims to tackle the problem of a lack of large datasets related to Broca's Aphasia.  
Patients of this neurological disorder lose the ability to form complete sentences and generate corresponding speech due to damage to the Broca's area in the brain. Large datasets of aphasic utterances are required to develop assistive devices for patients of this disorder using natural language processing models.

To solve this problem, we propose the use of a model that generates synthetic sentences that have linguistic features similar to aphasic utterances. We applied the proposed model to the C4 dataset comprising of simple, grammatically correct sentences from the internet. A subset of the C4 dataset of size 20,000 sentences had an average sentence of length 18.97 words. Overall, the noun-to-verb ratio in this subset was 2.96, meaning that it contains a significant number of sentences with two or three argument verbs. Moreover, for every such complex sentence, there are 1.70 simple sentences in the subset of this dataset. On the other hand, after applying the synthetic aphasic sentence generation model, the resultant dataset used for training with 20,000 sentences had an average length of utterance of 6.99 words and the testing dataset of 10,000 sentences had an average utterance of 7.01 words. Both the training and testing datasets had almost 9 times more simple sentences compared to complex sentences. 

The linguistic features of these synthetic datasets bear close similarity to statistical data collected from utterances of aphasia patients (\citet{thompson1995analysis, tetzloff2018quantitative}) where it was reported that aphasia patients speak sentences with an average length of 7.29 words. The average noun-to-verb ratio for aphasic speech was reported as 1.93, and the  percentage of utterance of complex sentences as almost negligible (\citet{tetzloff2018quantitative,thompson1995analysis}). This strong agreement of functional linguistic features of our synthetically generated aphasic speech with clinical statistical reports demonstrates the efficacy of our model.

Further, these datasets were used to tune the pre-trained T5 model to suggest five corrected sentences given an aphasic sentence. Upon the utterance of an agrammatic sentence by an aphasia patient, the model shall help the patient by providing choices of possibly intended sentences. The patient may then choose the best suited suggestion based on the actual intended sentence. The quality of suggestions made by the model was evaluated using scores that measure semantic similarity between two sentences.   The BLEU score is generally used to evaluate the performance of language translation models in which the sentence translated by a model is compared with the actual translation. For our predictive model, the intended sentence was compared with each of the suggested sentences. The pair with the highest score was considered to evaluate the model performance. With a BLEU score of 0.827 out of 1.000, our model showed effective completion of aphasic utterances in the testing dataset. Similar to the BLEU score, cosine semantic similarity was also used as a metric to quantify the performance of the model. Our model scored an average of 0.904 in semantic similarity out of 1.000 on the testing dataset. These results provide affirming foundation for the concept that a synthetic dataset based on small scale studies on aphasia can be used to develop effective assistive technology. The methodology proposed in this work can be extrapolated to study and analyse various other rare disorders involving brain injury or stroke.

\section{Conclusion}
In this study, we aim to put forth two contributions towards the research and development of assistive technology for patients of neurological disorders. First, we recognise that the research on assistive speech technology for patients suffering from Broca's aphasia is limited due to an unavailability of large scale studies on grammatical characterisation of affected speech. To overcome this deficit, a model for generating synthetic aphasic speech was proposed based on extensive study of small-scale linguistic analyses of speech production in Broca’s aphasia patients. A synthetic dataset was constructed from the C4 corpus using the proposed model. The linguistic features of the synthesized datsets were in agreement with the corresponding features reported in small-scale clinical studies on grammatical characterisation of aphasic speech.

Secondly, we demonstrate the utilization of the synthetic datasets for the development of assistive technology for patients of Broca's aphasia. The generated datasets were used to fine tune the T5 transformer to suggest correct sentences given an aphasic utterance. To evaluate the performance of the fine-tuned model, BLEU and semantic similarity scores were calculated between the target and suggested sentences. The obtained scores were found to be affirming using the tuned model. These results offer a proof of the concept that a synthetic dataset based on the linguistic features of agrammatic aphasia can be used as an effective proxy to clinically collected data of smaller sizes. The present study highlights our attempt to re-initiate research in this field and also bridge the research gap between neuroscience, linguistics, and natural language processing. Detailed quantitative characterisation of aphasic speech is required to build more accurate linguistic models. While disorders like Broca’s aphasia offer a small sample size of patients and data, synthetic linguistic models like ours offer extensive scope for developing assistive technology and rehabilitation monitoring.

\section{Challenges and Future Scope}
The study of Broca’s aphasia for neurological and linguistic analyses poses certain challenges.
The number of available subjects for such studies are very less.
Further studies are required regarding the neural pathways for language processing to effectively model the sentence production in the brain.
Observations regarding functional analysis of aphasic speech are based on smaller number of subjects. Generalised conclusions derived may not hold true for the disorder.
Linguistic analysis of aphasic speech in various languages shall converge for a generalised model of agrammatism and role of Broca’s area in language processing.

The present work aims to introduce a concept that can be expanded in numerous ways. Generation of synthetic datasets for language and speech disorders based on clinical studies may prove to be an effective method for rapid realisation of assistive technology. Moroever, such models may help us understand the disorders from a functional perspective. In the future, wider studies on linguistic analysis of affected speech may be done to develop a more accurate models for various disorders.







\bibliographystyle{elsarticle-num-names}
\bibliography{aphasia_refs.bib}







\end{document}